\documentclass[aps,prl,twocolumn,superscriptaddress]{revtex4}

\usepackage{graphicx}
\usepackage{color}

\begin{document}

\title{Ultrafast Electron Dynamics in Epitaxial Graphene Investigated with Time- and Angle-Resolved Photoemission Spectroscopy}

\author{S\o ren Ulstrup}
\affiliation{Department of Physics and Astronomy, Interdisciplinary Nanoscience Center (iNANO), Aarhus University, Denmark}
\author{Jens Christian Johannsen}
\affiliation{Institute of Condensed Matter Physics, \'Ecole Polytechnique F\'ed\'erale de Lausanne (EPFL), Switzerland}
\author{Alberto Crepaldi}
\author{Federico Cilento}
\affiliation{Sincrotrone Trieste, Trieste, Italy}
\author{Michele Zacchigna}
\affiliation{IOM-CNR Laboratorio TASC, Area Science Park,Trieste, Italy}
\author{Cephise Cacho}
\author{Richard T. Chapman}
\affiliation{Central Laser Facility, STFC Rutherford Appleton Laboratory, Harwell, United Kingdom}
\author{Emma Springate}
\affiliation{Central Laser Facility, STFC Rutherford Appleton Laboratory, Harwell, United Kingdom}
\author{Felix Fromm}
\affiliation{Lehrstuhl f{\"u}r Technische Physik, Universit{\"a}t Erlangen-N{\"u}rnberg,Germany}
\author{Christian Raidel}
\affiliation{Lehrstuhl f{\"u}r Technische Physik, Universit{\"a}t Erlangen-N{\"u}rnberg,Germany}
\author{Thomas Seyller}
\affiliation{Institut f{\"u}r Physik, Technische Universit{\"a}t Chemnitz , Germany}
\author{Fulvio Parmigiani}
\affiliation{Sincrotrone Trieste, Trieste, Italy}
\affiliation{Department of Physics, University of Trieste, Italy}
\author{Marco Grioni}
\affiliation{Institute of Condensed Matter Physics, \'Ecole Polytechnique F\'ed\'erale de Lausanne (EPFL), Switzerland}
\author{Philip Hofmann}
\affiliation{Department of Physics and Astronomy, Interdisciplinary Nanoscience Center (iNANO), Aarhus University, Denmark}
\email{philip@phys.au.dk}


\begin{abstract}
In order to exploit the intriguing optical properties of graphene it is essential to gain a better understanding of the light-matter interaction in the material on ultrashort timescales. Exciting the Dirac fermions with intense ultrafast laser pulses triggers a series of processes involving interactions between electrons, phonons and impurities. Here we study these interactions in epitaxial graphene supported on silicon carbide (semiconducting) and iridium (metallic) substrates using ultrafast time- and angle-resolved photoemission spectroscopy (TR-ARPES) based on high harmonic generation. For the semiconducting substrate we reveal a complex hot carrier dynamics that manifests itself in an elevated electronic temperature and an increase in linewidth of the $\pi$ band. By analyzing these effects we are able to disentangle electron relaxation channels in graphene. On the metal substrate this hot carrier dynamics is found to be severely perturbed by the presence of the metal, and we find that the electronic system is much harder to heat up than on the semiconductor due to screening of the laser field by the metal. 
\end{abstract}
\maketitle


The excitation of a crystal with an intense ultrafast light pulse brings the electrons of the material out of equilibrium. The equilibrium conditions are restored through a series of relaxation processes on different timescales including carrier scattering, hot carrier thermalization and finally thermal equilibrium with the lattice \cite{sundaraminducing2002}. Obtaining a consistent picture of such processes for graphene is not straightforward due to the peculiar low-energy spectrum consisting of a Dirac cone that is formed by the $\pi$ and $\pi^{\ast}$ bands crossing in proximity of the Fermi level, which places the optical properties of the material somewhere between those of metals and semiconductors \cite{Tse:2009,Malic:2011}. This intermediacy leads to a series of intriguing traits such as constant absorption \cite{Nair:2008} and the possibility of achieving carrier multiplication \cite{Winzer:2010,Song:2013,Tielrooij:2012,ploetzing:2014}. Combined with a high carrier mobility and gate tunable carrier doping, these form the basis of very attractive perspectives for photoelectric devices such as efficient solar cells or photodetectors \cite{Bonaccorso:2010, Mueller:2010aa,Gabor:2011}. Since electronic or optoelectronic devices that exploit these properties inevitably involve hot carrier transport, it is crucial to determine how hot carriers are generated and how they decay when graphene is excited by light. 

The light-matter interaction in graphene has been intensely studied using all-optical approaches \cite{Breusing:2011,Brida:2013aa,Shang:2011,Tielrooij:2012}, however in order to directly access the electron dynamics of the Dirac particles a technique that simultaneously offers energy-, momentum- and time-resolution is called for. Angle-resolved photoemission spectroscopy (ARPES) is the ideal method to measure the band structure and many-body effects of condensed matter materials. The advent of ultrafast intense laser sources and the development of high harmonic generation (HHG) of extreme ultraviolet (XUV) pulses have carried this technique into the time domain. By creating a transient distribution of electrons in the unoccupied bands of a material with an infrared pump pulse and subsequently performing ARPES with an XUV pulse, it is possible to simultaneously acquire spectral and dynamic information about the out-of-equilibrium carrier excitation and relaxation processes at arbitrary momenta in the Brillouin zone (BZ) \cite{Rohwer:2011}. Such time-resolved ARPES (TR-ARPES) measurements have been undertaken with near UV laser sources. The probing photon energy is however limited to $\sim6$~eV \cite{schmitt2008}, which is not enough to collect TR-ARPES data from electronic states with large in-plane momentum $k_{\|}$ due to the constraints imposed by energy and momentum conservation. For example, to photoemit electrons from the Dirac cone at $k_{\|} = 1.7$~\AA$^{-1}$ in graphene, photon energies of at least 16~eV are needed. With XUV-based TR-ARPES this has come within experimental reach as exemplified by recent TR-ARPES studies of monolayer graphene \cite{Johannsenb:2013,Gierz:2013aa,Someya:2014,gierznonequilibrium2014,Bignardi:2014}, bilayer graphene \cite{Ulstrup:2014,Gierz:2014ar} and graphite \cite{Stange:2013}.

\begin{figure*}[t!]
\includegraphics[width=1\textwidth]{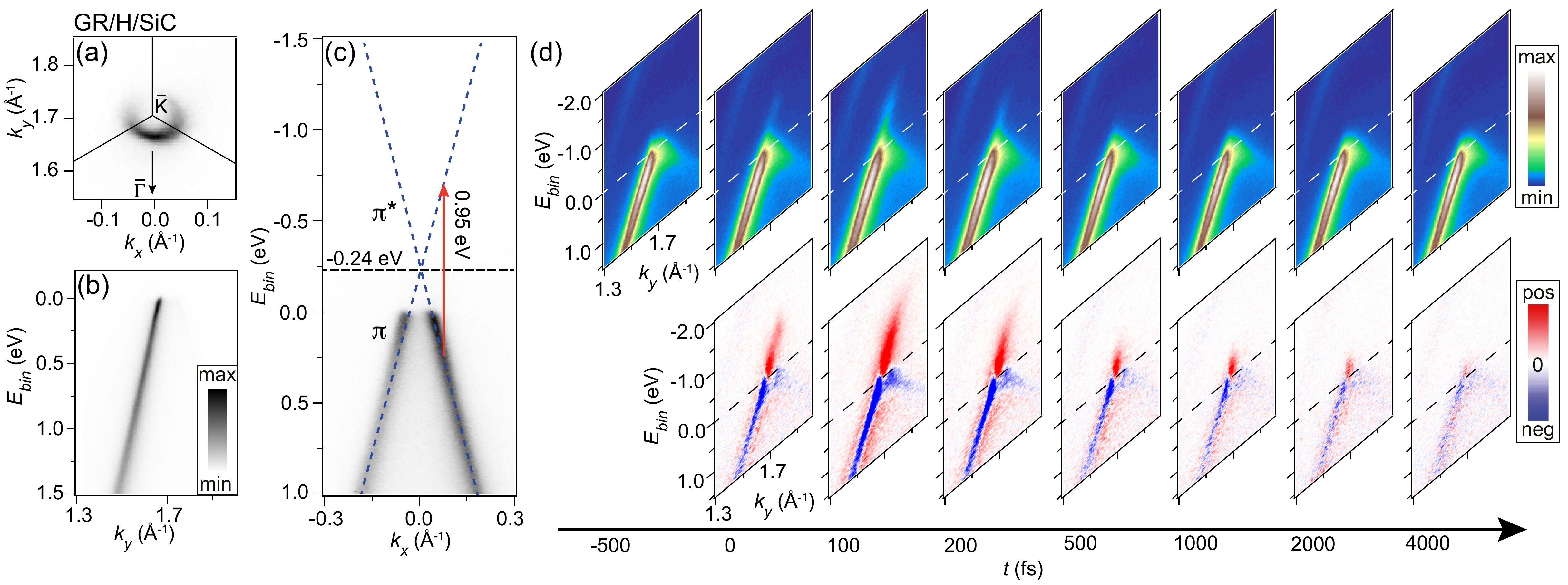}\\
\caption{Equilibrium and out-of-equilibrium photoemission measurements for GR/H/SiC: (a)-(c) ARPES data obtained with synchrotron radiation revealing (a) the Fermi contour of graphene around the BZ corner, (b) a single intense linear branch in the $\bar{\mathrm{\Gamma}}-\bar{\mathrm{K}}$ direction and (c) both branches along a line orthogonal to  $\bar{\mathrm{\Gamma}}-\bar{\mathrm{K}}$. In (c) the bands have been linearly extrapolated (dashed lines) to show the position of the Dirac point (horizontal dashed line) at -0.24~eV. (d) TR-ARPES spectra (top row) and difference spectra (bottom row) at the time delays marked on the timeline. The excitation is caused by a pump beam of 0.95~eV initially leading to vertical transitions as sketched by the red arrow in (c). The equilibrium Fermi level is marked by dashed lines in (d).}
\label{fig:1}
\end{figure*}

Here we discuss the hot carrier dynamics in epitaxial graphene samples grown on both semiconducting and metallic substrates on the basis of TR-ARPES measurements. The response to laser excitations for both types of substrates are important for the operation of actual photoelectric devices as semiconducting substrates are employed for gating and metallic contact pads will always be in close proximity of the graphene sheet \cite{Bonaccorso:2010}. To study the implications of these substrates for the electron dynamics around the Dirac cone, so-called quasi-free standing graphene samples are employed as they exhibit an unperturbed Dirac dispersion. Specifically, we use hydrogen intercalated graphene on 6H-SiC(0001) (GR/H/SiC) \cite{Riedl:2009,Speck:2011} and oxygen intercalated graphene on Ir(111) (GR/O/Ir) \cite{Larciprete:2012,ulstrupseq:2014}.  Both samples are hole-doped but with different carrier concentrations. Parts of the results for GR/H/SiC have been published previously \cite{Johannsenb:2013} but are reviewed here to make the new data and the comparison to GR/O/Ir more accessible.  

The TR-ARPES measurements on both samples were performed at the ARPES end-station at the Artemis facility, Rutherford Appleton Laboratory \cite{Cacho:2014,Frassetto:2011}. Pump and probe pulses were generated by a 1~kHz Ti:sapphire amplified laser system, which provided infrared pulses centered at a wavelength of 785~nm with a duration of 30~fs and an energy of 10 mJ per pulse. A $p$-polarized probe beam with a photon energy of 33.2~eV, corresponding to the 21st harmonic of the HHG spectrum generated in a pulsed Ar gas jet synchronized with the drive laser, was selected using a time-preserving monochromator. Tunable pump pulses were obtained by seeding an optical parametric amplifier (HE-Topas), and 1.6 eV  energy photons were generated by higher harmonics of the HE-Topas light. The pump beam was $s$-polarized to avoid any interference between the laser field and the photoemitted electrons \cite{crepaldi2013}. The samples were kept at a temperature of 300~K, and the total time, energy and angular resolution were set to 60~fs, 350~meV and~0.3$^{\circ}$, respectively. In order to map the equilibrium electronic structure of the samples with high resolution, synchrotron-based ARPES measurements were performed with a photon energy of 47~eV at the SGM-3 beamline on the ASTRID light source in Aarhus, Denmark \cite{Hoffmann:2004}. The energy- and angular-resolution in these measurements were 18~meV and 0.1$^{\circ}$, respectively, and the samples were kept at a temperature of 70~K using a He cryostat.

GR/H/SiC is produced \emph{ex situ} by thermal decomposition of the Si-terminated face of SiC(0001) in an argon atmosphere until a so-called buffer layer is formed. Intercalation of hydrogen then decouples the buffer layer from the substrate, which yields the pristine quasi-free standing graphene sheet with the Dirac cone clearly defined in the synchrotron ARPES measurements in Fig. 1(a)-(c) \cite{Riedl:2009,Speck:2011}. This quasi-free-standing monolayer graphene is $p$-doped with a carrier concentration of $\approx$ 4 $\times 10^{12}$ cm$^{-2}$, that places the Dirac point 240 meV above the Fermi energy as seen by the linear extrapolation of the bands in Fig. 1(c). The TR-ARPES data are acquired along the $\bar{\mathrm{\Gamma}}-\bar{\mathrm{K}}$ direction marked by the arrow on the Fermi contour map in Fig. 1(a). Along $\bar{\mathrm{\Gamma}}-\bar{\mathrm{K}}$ the intensity peaks in the linear branch in the first BZ while it is completely extinguished in the second BZ due to well-known matrix element effects \cite{Shirley:1995b,MuchaKruczynski:2008}, which leads to the single intense branch in Fig. 1(b). Both branches are visible when scanning along a direction orthogonal to $\bar{\mathrm{\Gamma}}-\bar{\mathrm{K}}$ as seen in  Fig. 1(c). 

The TR-ARPES measurements are performed along $\bar{\mathrm{\Gamma}}-\bar{\mathrm{K}}$ for a pump excitation of $\hbar\omega = 0.95$~eV with a pump fluence of 340 $\mu$J/cm$^2$ that induces vertical transitions between the $\pi$ and $\pi^{\ast}$ bands as sketched in Fig. 1(c). Snapshots of the dispersion are acquired for time steps of 100~fs both before (negative time delays) and after (positive time delays) the arrival of the optical excitation, as shown in Fig. 1(d). As the position of the chemical potential varies after optical excitation \cite{Crepaldi:2012}, the binding energy scale refers to the position of the Fermi level at negative time delays, where the system is in equilibrium. The top panels of Fig. 1(d) display the photoemission intensity as a function of the delay time, while the bottom panels display the difference spectra obtained by subtracting the photoemission intensity signal at negative time delays from the corresponding spectra at positive time delays. Contrary to the synchrotron ARPES data, the second branch of the $\pi$ band is slightly visible, which may be caused by the poorer angular resolution leading to photoemission intensity from a broader range of angular distributions. A duplicate of the most intense branch is also visible in the binding energy range from $E_{bin}=-2$~eV to $E_{bin}=-1$~eV, which is caused by a small amount of photons associated to the 23rd harmonic in the XUV pulse.

\begin{figure} [t!]
\includegraphics[width=0.5\textwidth]{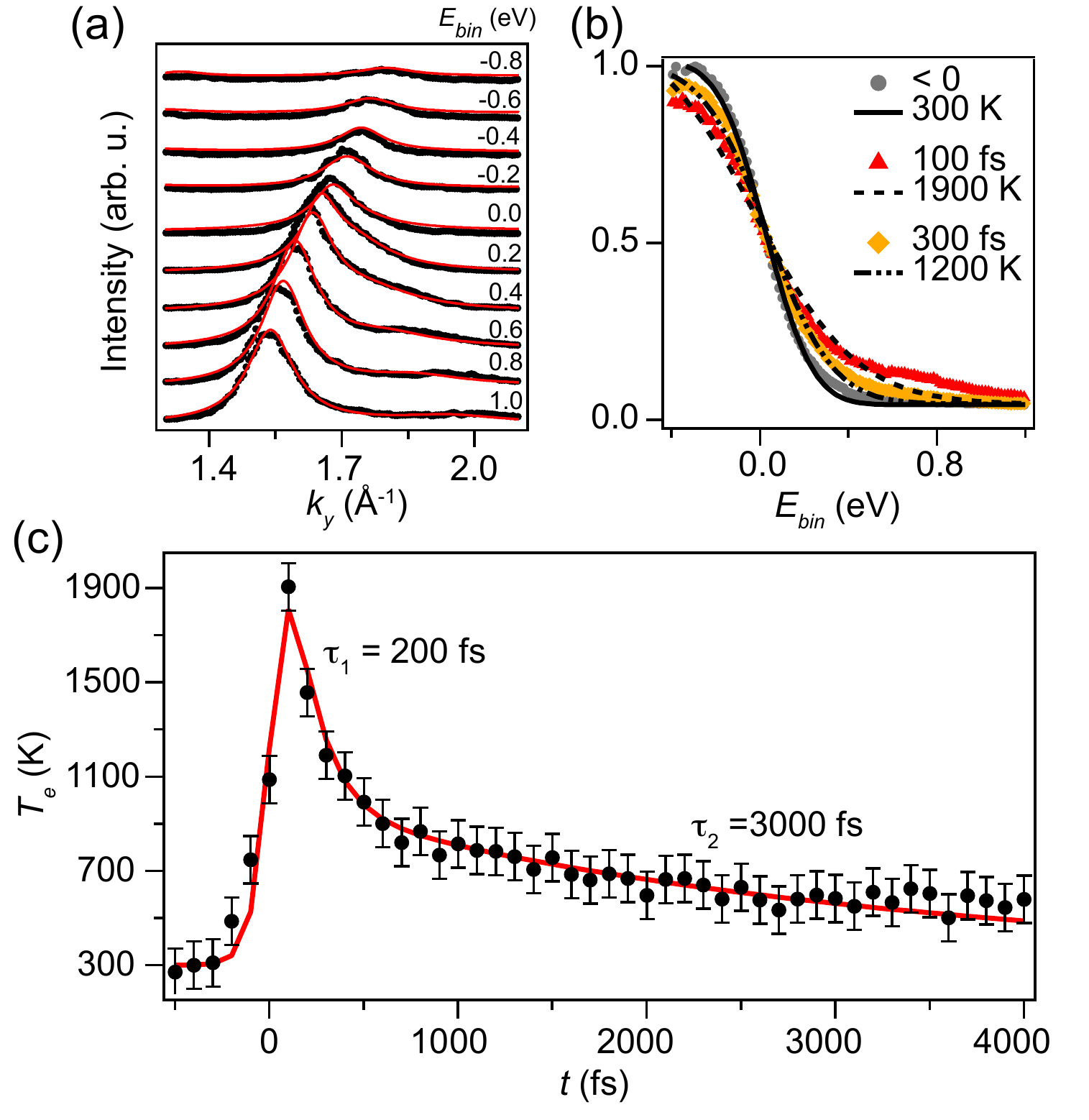}
\caption{Extraction of the electronic temperature from an analysis of momentum distribution curves (MDCs): (a) Subset of MDCs (markers) fitted by double Lorentzian functions (lines) for the TR-ARPES spectrum at a time delay of 100~fs. (b) Integrated Lorentzians as a function of binding energy (markers) at the given time delays. The fits consist of Fermi-Dirac functions broadened with Gaussians that take the energy resolution into account (lines), which provide the stated electronic temperatures. (c) Fitted electronic temperature as a function of time delay (markers). The line corresponds to a fitting function that describes the rising edge and two exponential decays with the given time constants $\tau_1$ and $\tau_2$, convoluted with a Gaussian that takes the time resolution into account.}
\label{fig:2}
\end{figure}

To a first approximation, the excess intensity (red) above $E_{bin}=0$ in the difference spectra can be viewed as the electrons injected by the pump, while the depletion of intensity below $E_{bin}=0$ is due to the corresponding holes. 
Within the optical excitation at $t = 0$, the excess intensity increases. This signifies a pumping of electrons (holes) into the $\pi^{\ast}$ ($\pi$) band. Already during the optical excitation the distribution of holes and electrons appears uniform along the branch i.e. we do not observe the filling (emptying) of the discrete states in the $\pi^{\ast}$ ($\pi$) band  at $\hbar\omega/2$ above (below) the Dirac point. In fact, excited electrons are observed at energies substantially larger than the expected transition, which points towards Auger-like carrier scattering processes. A thermal distribution of hot electrons is thus established already within our time resolution. After the peak signal at $t = 100$~fs the excess intensity decays. Within the first 200~fs the removal of electrons appears efficient from regions of the branch at large negative binding energies while excited carriers in close vicinity of the Dirac point relax slowly on a timescale larger than 1~ps.

The time dependent temperature of electrons can be extracted from the photoemission intensity using a method involving the momentum distribution curves (MDCs) of the spectra at each time delay \cite{ulstrupc:2014}. This is illustrated for the spectrum at $t = 100$~fs in Fig. 2(a), where selected MDCs are shown along with the result of fits composed by two Lorentzian peaks (one for each branch) and a fixed parabolic background. Integrating the main Lorentzian peak and plotting this integral as a function of binding energy yields the energy- and time-dependent occupation function as shown in Fig. 2(b) for three time delays. Fitting these to Fermi-Dirac (FD) distributions, convoluted with a Gaussian function accounting for the finite energy resolution, yields the time dependence of the electronic temperature, which is shown in Fig. 2(c). Note that at $t=100$~fs additional intensity is seen in the experimental data above the FD fit in Fig. 2(b), which may be a hint of a non-thermal effect. However, this is difficult to conclude because an asymmetry in the matrix elements of the $\pi$ and $\pi^{\ast}$ bands can affect the shape of this curve \cite{MuchaKruczynski:2008,Ulstrup:2014,Gierz:2014ar}, and so can a slight misalignment of the cut through the Dirac point. We can rule out that the effect is due to separate distributions of electrons and holes as observed by Gierz \emph{et al.} \cite{Gierz:2013aa} since the threshold fluence for achieving this is 2~mJ/cm$^2$ \cite{Li:2012b}, and we are well below this regime at 340 $\mu$J/cm$^2$. We therefore assume that a thermalized hot electron distribution is present for all time delays after the optical excitation. Fitting the electronic temperature as a function of time delay with a
function that consists of an exponential rising edge and two exponential decays with time constants $\tau_1$ and $\tau_2$, convoluted with a Gaussian that takes the time resolution into account, reveals that the hot electron relaxation consists of a fast $\tau_1 = 200$~fs decay and a slow  $\tau_2 = 3000$~fs decay as shown in Fig. 2(c).

\begin{figure} [t!]
\includegraphics[width=0.5\textwidth]{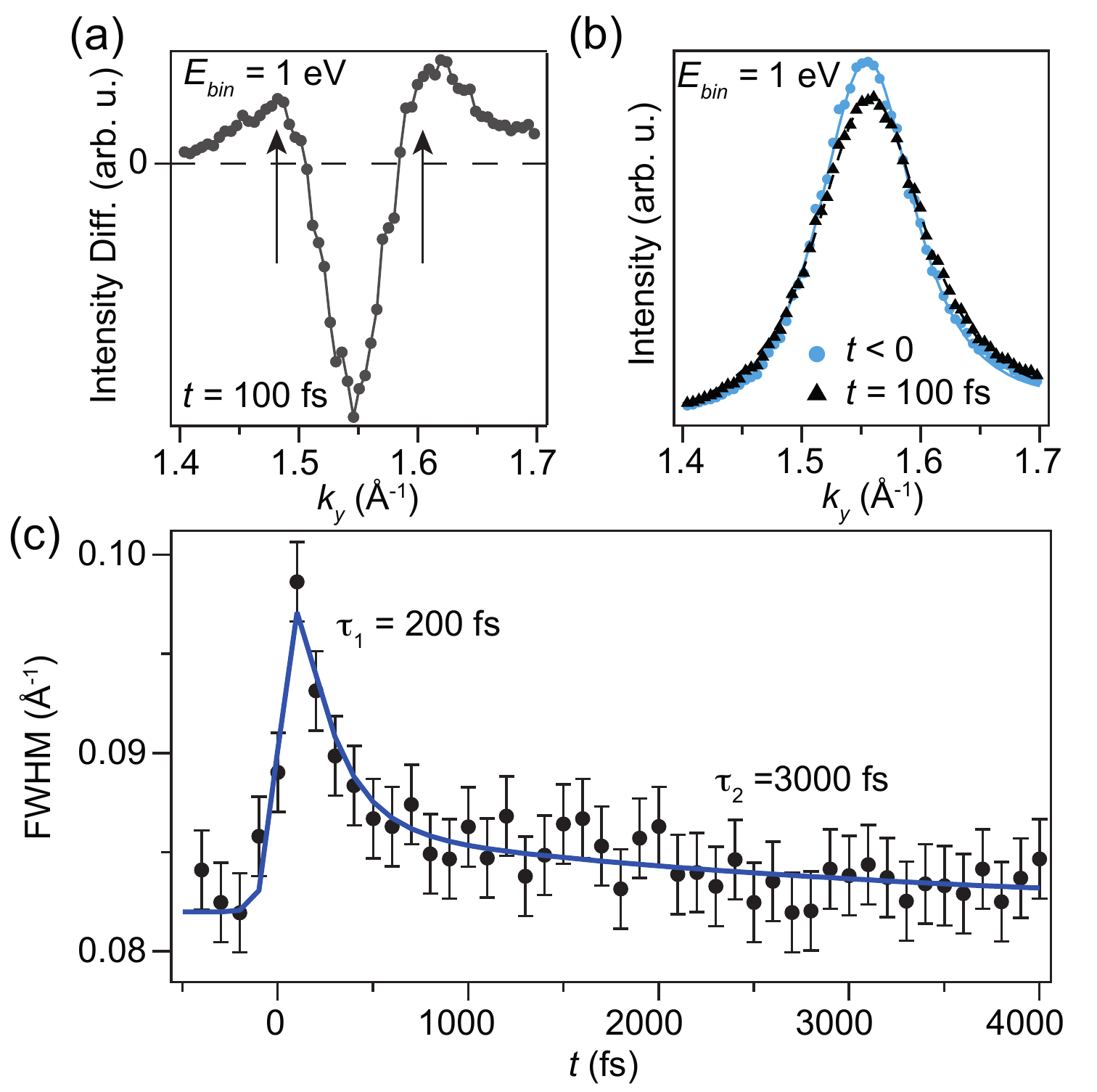}\\
\caption{Analysis of the time dependence of the $\pi$-band's linewidth:  (a) MDC of the difference spectrum at a binding energy of 1~eV, showing tails (marked by arrows) of excess intensity away from the main peak. The line is a guide to the eye. (b) MDCs (markers) at a binding energy of 1~eV binned over $\pm 0.1$~eV before and immediately after the arrival of the pump. Fits are shown by lines. (c) Full-width at half maximum (FWHM) of the MDC fits as a function of time delay. The line is a fitting function, similar to the one used for the electronic temperature in Fig. 2(c), where the decaying part is described by two exponential decays with the stated time constants $\tau_1$ and $\tau_2$.}
\label{fig:3}
\end{figure}

The optical excitation of the carriers does not only lead to their re-distribution, the spectral function is also affected, as seen in the changes of the $\pi$ band's linewidth. Such a change can be directly discerned in the difference spectra in the bottom row of Fig. 1(d). Apart from the obvious loss of intensity in the centre of the band (blue), the intensity distribution broadens in momentum and the extent of this effect appears to follow the timescales of the hot electron signal at negative binding energies. An MDC through the difference spectrum at $t=100$~fs and a binding energy of 1 eV is shown in Fig. 3(a), where the tails of excess intensity are clearly seen away from the main depression of intensity due to the pump-induced holes. To study this further an MDC through the photoemission intensity at a binding energy of 1~eV binned over  $\pm 0.1$~eV is fitted to a Lorentzian peak convoluted with a Gaussian that takes the momentum resolution of the experiment into account. This is done for all spectra in our time series. Examples are given for a spectrum before the arrival of the pump pulse and for the peak excited signal at $t=100$~fs in Fig. 3(b). The change of linewidth is tracked via the full-width at half maximum (FWHM) of the Lorentzian fits as a function of time delay as shown in Fig. 3(c). The dynamics of the linewidth is seen to duplicate the dynamics of the hot electron temperature with a double exponential fit consisting of 200~fs and 3000~fs decay constants describing the data. We can rule out that the broadening of the spectra is a space charge effect caused by a cloud of photoemitted secondary electrons since we do not observe any rigid shift in kinetic energy of the spectra when the pump arrives at the applied fluence of 340 $\mu$J/cm$^2$. If we increase the pump fluence above 1~mJ/cm$^2$ we observe such shifts, which are known to be caused by space charge \cite{hellmann2009}.  Furthermore, the broadening we observe is on a timescale of picoseconds while space charge effects persist over nanoseconds \cite{yangelectron2013}. 

\begin{figure*}[t!]
\includegraphics[width=1.0\textwidth]{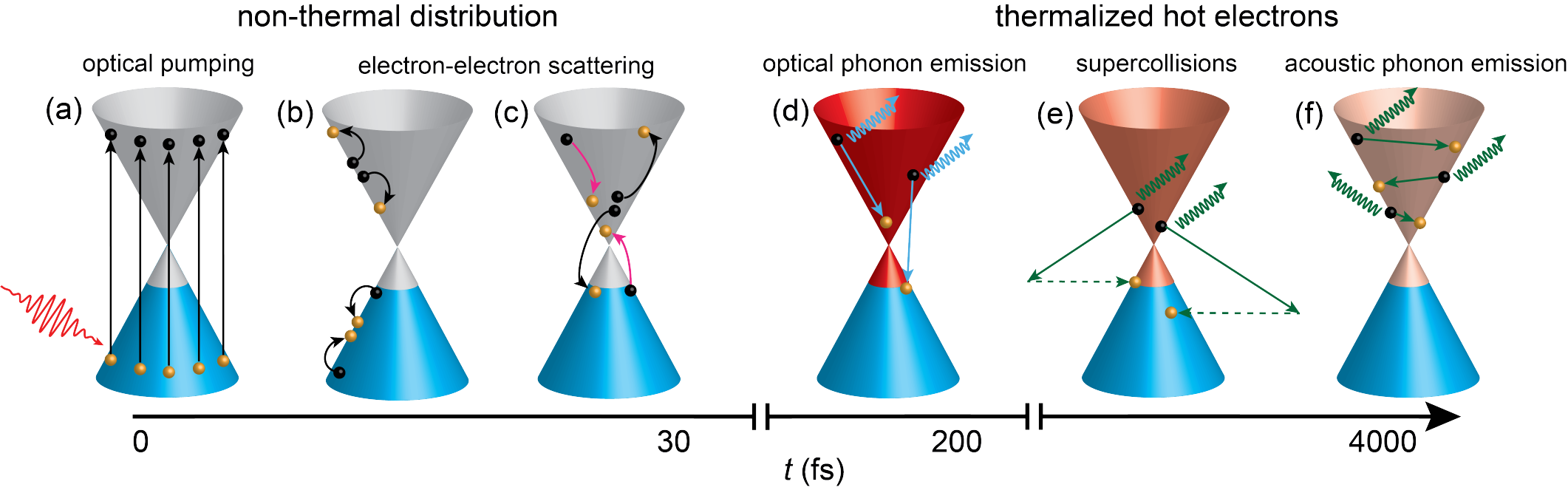}
\caption{Timeline of electron dynamics on the Dirac cone in graphene: (a) At time zero electrons (black spheres) and holes (yellow spheres) are generated by an infrared pump via vertical transitions (straight arrows). (b)-(c) Within 30~fs the excited electrons and holes undergo scattering processes (b) within and (c) between the bands by Auger recombination (black curled arrows) and impact excitations (magenta arrows). (d)-(f) This leads to a thermalized hot electron distribution that decays efficiently within 200~fs by (d) emission of optical phonons (blue wiggled arrows), which is followed by (e) slower supercollisions involving acoustic phonons (green wiggled arrows) and impurities (green dashed arrows). (f) Direct acoustic phonon emission also occurs, but is very inefficient.}
\label{fig:4}
\end{figure*}  
 
The dynamics of the hot electrons and the holes can be understood from the diagram in Fig. 4. We merely discuss the electrons here but equivalent arguments can be made for the holes. Initially, the pump pulse creates a non-thermal population of excited electrons in the $\pi^{\ast}$ band by direct transitions as shown in Fig. 4(a). These excited carriers thermalize via the electron-electron interaction on a time-scale of $\sim30$~fs \cite{Kampfrath:2005,Breusing:2009,Lui:2010,Brida:2013aa}, which is faster than the time resolution of our experiment. This leaves the electronic system at a well-defined temperature, which in our case reaches a maximum of 1900~K. Auger-like processes such as the intraband scattering of electrons and holes depicted in Fig. 4(b) as well as interband and intervalley processes between the two $\bar{\mathrm{K}}$ and $\bar{\mathrm{K}}^{\prime}$ points (not shown) all contribute to this fast thermalization \cite{Malic:2011}. The electron thermalization can be accompanied by carrier multiplication, i.e. the generation of more than one electron-hole pair per absorbed photon if the rate of impact ionization processes exceeds that of the inverse processes of Auger recombination. These interband processes are sketched in Fig. 4(c) \cite{Winzer:2010,Song:2011d}. The carrier multiplication effect is a function of fluence, time and doping \cite{Winzer:2012,Johannsen:2014}. These non-thermal processes are so fast that we are not able to directly observe them with the time resolution of our experiment.

Upon thermalization the hot electronic system can efficiently lose energy by emission of optical phonons. Intraband transitions by optical phonons cause the electrons to cascade down to the region close to the Dirac point while interband transitions mediate electron-hole recombination across the Dirac point. Both processes are sketched in Fig. 4(d). In addition to these cooling channels, intervalley scattering by zone boundary phonons also contribute to the cooling (not sketched) \cite{Butscher:2007}. These cooling channels are blocked for electrons with energies below the typical optical phonon energies in the range of $150-200$~meV or when the temperature of the optical phonons reaches that of the electrons \cite{Tse:2009,Wang:2010a,Tielrooij:2012}. Our data suggest that this occurs within the initial 200~fs through the initial fast decay of the electronic temperature.

A slow decay via acoustic phonons then sets in, which is responsible for the hot carrier relaxation time constant of 3000~fs. Due to the very small energy transfer involved in intraband acoustic phonon processes as illustrated in Fig. 4(f), such processes have to occur a multitude of times to bring the hot electron into the $\pi$ band. Roughly, the energy transfer in these processes scales as the ratio $v_s/v_F$, where $v_s = 2\cdot 10^{4}$~m/s is the sound velocity and $v_F = 10^{6}$~m/s is the Fermi velocity in graphene. The acoustic phonon mediated decay is therefore so inefficient that the relaxation would take several hundreds of picoseconds \cite{Bistritzer:2009}. Instead three-body collisions with defects and acoustic phonons can become the dominant contribution, bringing the relaxation time down to the order of a few picoseconds \cite{Song:2012e,Betz:2012,Graham:2012}. These so-called supercollisions are especially relevant for carriers near the Dirac point region as shown in Fig. 4(e), and mediate a faster relaxation across the Dirac point that would otherwise impose a bottleneck for acoustic phonon cooling. This picture of the phonon cooling of hot carriers in graphene is substantiated by a phenomenological three-temperature model as described in our previous study \cite{Johannsenb:2013}.

The time-dependent broadening of the $\pi$ band seen in Fig. 3(c) reflects the initial pump-induced opening of additional decay channels for the holes as sketched in the $\pi$ band part in Fig. 4(b) \cite{Rana:2007}. After the peak signal at $t=100$~fs these additional channels are closed by the optical and acoustic phonon processes that mediate electron-hole recombination between the $\pi$ and $\pi^{\ast}$ bands. This interpretation is supported by the strong similarity with the time dependence of the hot electron temperature, which is essentially caused by the same processes.

\begin{figure*}[t!]
\includegraphics[width=1.0\textwidth]{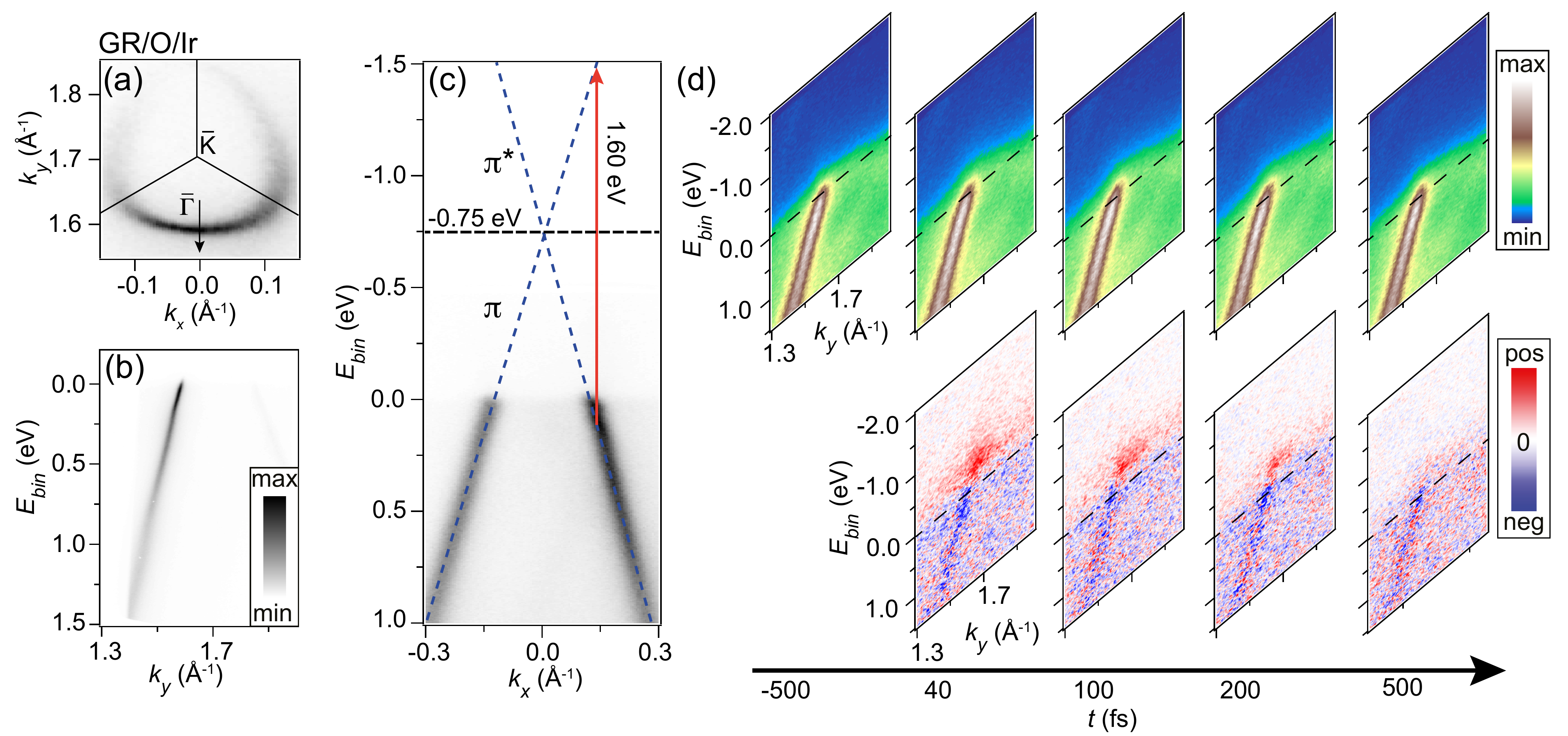}
\caption{Equilibrium and out-of-equilibrium photoemission measurements for GR/O/Ir: (a)-(c) Synchrotron ARPES measurements of (a) Fermi contour, and $\pi$ bands in (b) the $\bar{\mathrm{\Gamma}}-\bar{\mathrm{K}}$ direction and (c) orthogonal to this direction. The linear extrapolation (dashed lines) of the bands in (c) reveal the Dirac point position (horizontal dashed line) at -0.75~eV. (d) TR-ARPES spectra (top row) and difference spectra (bottom row) at the time delays marked on the timeline. The pump energy is tuned to $\hbar\omega = 1.60$~eV enabling the vertical transitions sketched by the red arrow in (c). The equilibrium Fermi level is marked by dashed lines in (d).}
\label{fig:5}
\end{figure*} 

We now turn to the GR/O/Ir case, which permits an investigation of the effect of larger hole doping and, more importantly, the implications of a metal substrate for a laser TR-ARPES experiment. These samples are prepared on clean single crystal Ir(111) substrates by temperature-programmed growth cycles, where ethylene gas at a background pressure of $5\times10^{-7}$~mbar is dosed in an ultra high vacuum chamber while ramping the temperature of the Ir crystal between 520~K and 1520~K. This procedure routinely yields single crystal monolayer graphene of high quality \cite{Pletikosic:2009,Kralj:2011}. Oxygen is then intercalated by applying a high oxygen pressure and elevated temperatures. These samples are generally strongly hole doped with a doping level that can be slightly tuned by the amount of intercalated oxygen \cite{Larciprete:2012}. In this study we employ a sample with a carrier concentration of 4 $\times 10^{13}$ cm$^{-2}$, placing the Dirac point 750 meV above the Fermi energy as shown in the synchrotron ARPES data in Figs. 5(a)-(c). The Fermi contour is therefore significantly larger than in the GR/H/SiC case in Fig. 1(a). A single sharp and quasi-free standing branch is seen in the $\bar{\mathrm{\Gamma}}-\bar{\mathrm{K}}$ direction in Fig. 5(b) and both branches are discerned orthogonal to this direction in Fig. 5(c).

Since GR/O/Ir exhibits a larger hole doping than GR/H/SiC it is necessary to tune the pump energy to $\hbar\omega = 1.60$~eV in order to enable direct transitions between the $\pi$ and $\pi^{\ast}$ bands as sketched in Fig. 5(c). In the present experiment a substantially larger pump fluence has been used, 31 mJ/cm$^2$, owing to the weaker response of the material.  A time step of 40~fs is used to acquire these data. This yields the spectra in Fig. 5(d). Compared to Fig. 1(d) a stronger background is observed in addition to the main $\pi$ band branch, which is caused by the underlying metal substrate \cite{ulstrupseq:2014}. The difference spectra in the bottom row of Fig. 5(d) show that the excess intensity decays fast, and that it appears localized in close vicinity of the equilibrium Fermi level. 

The fast decay is studied further in Fig. 6, where difference spectra are shown immediately after the pumping stage at $t=40$~fs and at a later stage at $t=400$~fs. The observed response at 40~fs in Fig. 6(a) to the large applied fluence appears to be extremely weak, and a surplus of intensity is observed for only a small range of binding energies above the equilibrium Fermi level. This observation clearly points towards a substantial screening of the electric field of the pump light by the underlying metal, and consequently a highly reduced population of excited carriers in the $\pi$ band branches. Since the pump beam is $s$-polarized with respect to the scattering plane it is oriented parallel to the surface. Within a classical dipole approximation this field is cancelled out at the surface of the metal. This description is valid only if the plasma frequency $\omega_p$ of the metal is substantially larger than the pump field frequency $\omega$, which is fulfilled in our case since for metals $\hbar\omega_p$ is in the range of $6-15$~eV and the applied pump energy is $\hbar\omega \sim 1.60$~eV \cite{Zangwill:1988}. The timescale of this screening is determined by the plasma frequency and therefore below 1~fs, which is much faster than our experiment. Population of the $\pi^{\ast}$ bands at $\hbar\omega/2$ above the Dirac point is not at any point in time observed in our data. Furthermore, contrary to the GR/H/SiC case we do not even observe pumping of the $\pi^{\ast}$ band. The excited carriers in GR/O/Ir decay back to the equilibrium state on a time scale of less than 400~fs as indicated by the difference spectrum in Fig. 6(b). This suggests that the lifetime of the hot carriers is reduced in graphene supported by a metal surface. The observed significant surplus of intensity above the Fermi level in Fig. 6(a), even for $k_{||}$ away from the graphene dispersion could help to explain this ultrafast decay of the excited carriers in graphene. It seems reasonable to assume that this background intensity corresponds to carriers excited in the metal. In Fig. 6(c), we compare the dynamics of the background intensity to the intensity corresponding to the excited carriers in graphene by integrating the intensity in the two boxed regions shown in Figs. 6(a)-(b). For both regions, the integrated intensity is shown to follow a single exponential decay with the same decay constant of $\approx$ 250~fs. The observed identical dynamics in the two regions may be due to the fact that the relaxation dynamics of excited carriers in graphene on this metal substrate involve a decay through the states of the metal. However, given that the graphene electronic structure is well decoupled from the substrate in this system \cite{Larciprete:2012,ulstrupseq:2014} it seems more likely that the photoinduced hot electron signal in the $\pi$ band is weakened compared to the GR/H/SiC case due to a significant screening of the laser field. Note that the field is not entirely screened since a weak optical pumping is observed in the $\pi$ band in Fig. 6(a).

\begin{figure} [t!]
\includegraphics[width=0.5\textwidth]{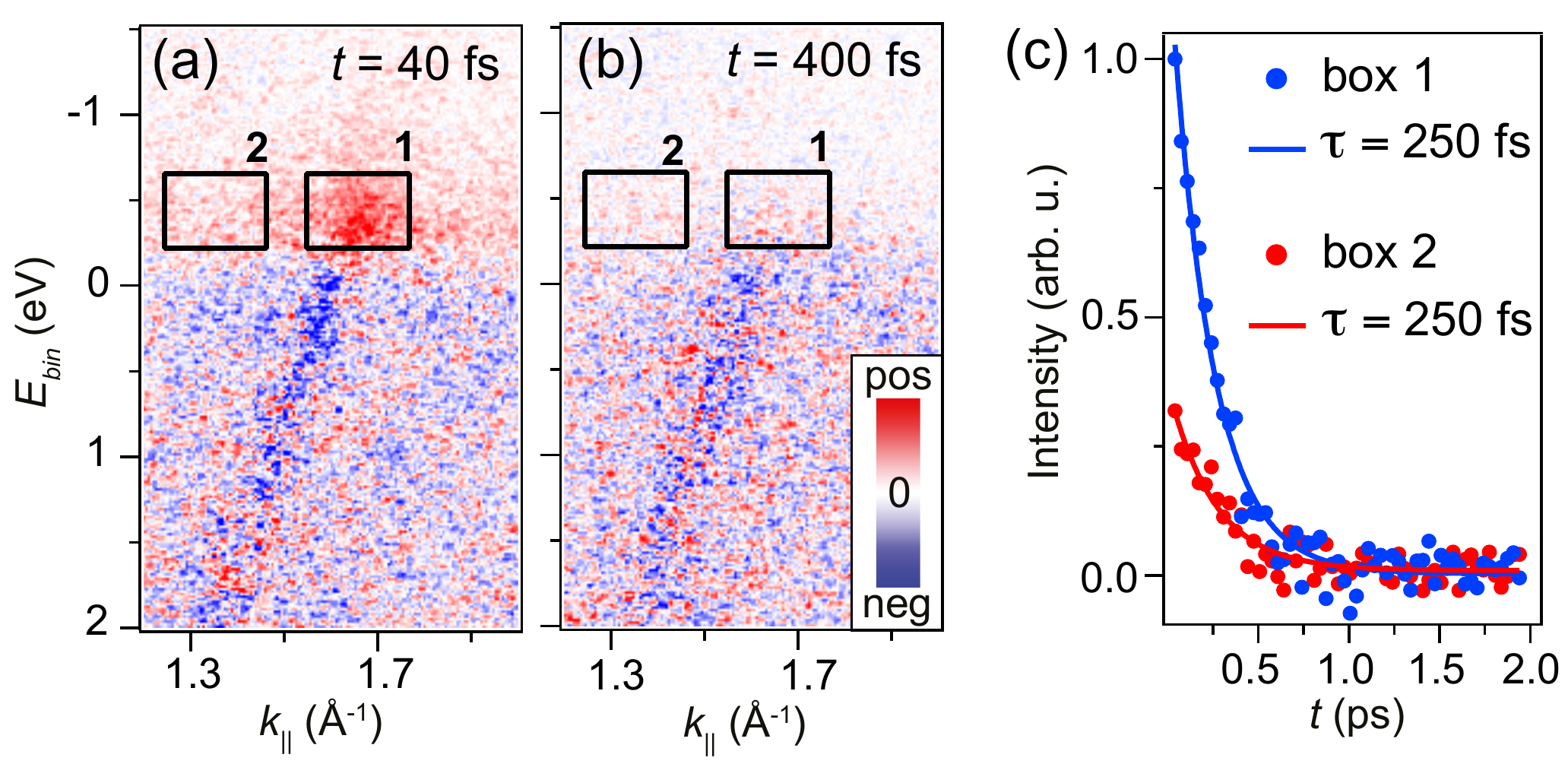}\\
\caption{Dynamics in GR/O/Ir: (a)-(b) difference spectra at (a) 40~fs and (b) 400~fs. (c) Intensity integrated in the boxed regions shown in (a)-(b) and normalized to the maximum intensity in box 1. Lines are single exponential fits with a decay time of $\tau = 250$~fs in both cases.}
\label{fig:6}
\end{figure}

In conclusion, using TR-ARPES on quasi-free standing graphene on a semiconducting SiC substrate it is possible to induce excitations from the $\pi$ to the $\pi^{\ast}$ band and observe dynamics of hot carriers. These thermalize on an ultrashort timescale that is shorter than the experimental time resolution of 60~fs. The hot carrier dynamics is dominated by a fast 200~fs optical phonon mediated decay and a slow 3000~fs decay due to acoustic phonons and impurities (supercollisions). For a more detailed analysis of this, see Ref. \cite{Johannsenb:2013}. This dynamics is seen in the hot electron temperature, which is extracted by analyzing the optically induced transient population of the $\pi^{\ast}$ band, and it is also reflected in a time-dependent broadening of the $\pi$ band due to the time-dependence of the available decay channels for photoinduced holes. We find that for graphene on a metal substrate, here oxygen intercalated graphene on Ir(111), it is not possible to observe such hot carrier dynamics due to the metal screening the laser field. This is an important result for laser-based experiments as metals are expected to severely influence the measurement, and it may play a role for graphene based photoelectric devices where metallic contact pads can affect the graphene-light interaction. 

We gratefully acknowledge financial support from the VILLUM foundation, The Danish Council for Independent Research / Technology and Production Sciences, the Swiss National Science Foundation (NSF), EPSRC, The Royal Society and the Italian Ministry of University and Research (Grants No. FIRBRBAP045JF2 and No. FIRB-RBAP06AWK3). Access to the Artemis Facility was funded by STFC. Work in Chemnitz was supported by the European Union through the project ConceptGraphene, and by the German Research Foundation in the framework of the SPP 1459 Graphene.
 




\end{document}